# BHT-QAOA: Generalizing Quantum Approximate Optimization Algorithm to Solve Arbitrary Boolean Problems as Hamiltonians

Ali Al-Bayaty

*Electrical and Computer Eng. Dept.*

*Portland State University*

*Portland, USA*

albayaty@pdx.edu

orcid.org/0000-0003-2719-0759

Marek Perkowski

*Electrical and Computer Eng. Dept.*

*Portland State University*

*Portland, USA*

h8mp@pdx.edu

orcid.org/0000-0002-0358-1176

***Abstract*–**A new methodology is proposed to solve classical Boolean problems as Hamiltonians, using the quantum approximate optimization algorithm (QAOA). Our methodology successfully finds all optimized approximated solutions for Boolean problems, after converting them from Boolean oracles (in different structures) into Phase oracles, and then into the Hamiltonians of QAOA. From such a conversion, we noticed that the total utilized numbers of qubits and quantum gates are dramatically minimized for the final quantum circuits of Hamiltonians. In this paper, arbitrary classical Boolean problems are examined by successfully solving them with our proposed methodology, using structures based on various logic synthesis methods, an IBM quantum computer, and a classical optimization minimizer. Accordingly, this methodology will provide broad opportunities to solve many classical Boolean problems as Hamiltonians, for the practical engineering applications of several algorithms, robotics, machine learning, just to name a few, in the hybrid classical-quantum domain.

**Keywords:** *Boolean oracles, Phase oracles, logical structures, logic synthesis, Hamiltonians, QAOA*

## 1 Introduction

The quantum approximate optimization algorithm (QAOA) was introduced by Farhi *et al.* [1, 2] to solve combinatorial optimization problems, such as the MaxCut problem [3, 4], in the quantum domain. The MaxCut problem is a subset of the classical graph theoretical problems, which is represented by a number of nodes ($n$) connected through a set of edges {$j$, $k$}, where $j$ and $k$ are the indices of two connected nodes. The solutions for a MaxCut problem can be determined by finding the maximum number of cuts in edges for the connected nodes. In the quantum domain, the QAOA represents the MaxCut problem in the form of ansatz Hamiltonian oracle, which is termed the "Hamiltonian clauses ($H_C$)", and ansatz Hamiltonian operator, which is termed the "Hamiltonian mixer ($H_M$)". The wording "ansatz" means that $H_C$ and $H_M$ consist of parameterized rotational Pauli-Z and Pauli-X gates, respectively. Such that $H_C$ consists of a number of RZ($v·γ$), RZZ($v·γ$), RZZZ($v·γ$), and so on, while $H_M$ consists of $n$ numbers of RX($ω·β$), where $v$ and $ω$ are the coefficients (as the time evolutions for both Hamiltonians), $γ$ and $β$ are the parameterized angular rotations, and $n$ is the number of input qubits. Note that $γ$ and $β$ are restrictively rotated between [0, 2π] and [0, π], respectively [1, 2].

In general, the following steps illustrate the complete construction of QAOA, to find all solutions to a classical MaxCut problem:

1. All $n$ nodes of the MaxCut problem are represented into their equivalent $n$ input qubits, which are initially set to the $|0\rangle$ state.
2. Hadamard (H) gates are applied to all $n$ input qubits, to create the complete quantum search space of $\{0,1\}^{\otimes n}$ for QAOA to find all solutions.
3. Hamiltonian $H_C$ represents the quantum circuit of a MaxCut problem as the unitary operator $\left(e^{-i\gamma H_C}\right)$, which is a set of non-connected nodes (as $RZ_j(v \cdot \gamma)$ gates) and connected nodes (as $RZ_jZ_k(v \cdot \gamma)$ gates), where $j$ and $k$ are the indices of input qubits (as nodes).
4. Hamiltonian $H_M$ represents the quantum circuit for the sum of all $n$ input qubits as the unitary operator $\left(e^{-i\beta H_M}\right)$, which is a set of $n$ $RX(\omega \cdot \beta)$ gates. Note that $H_M$ acts as the diffusion operator of QAOA analogous to the diffusion operator in Grover's algorithm [5-8].
5. To improve the quality of all approximated solutions, $H_C$ and $H_M$ are iterated for a number of repetitions ($p$), where $p \geq 1$. Such that every $e^{-i\gamma_p H_C}$ consists of $RZ(v \cdot \gamma_p)$, $RZZ(v \cdot \gamma_p)$, etc., and every $e^{-i\beta_p H_M}$ consists of $RX(\omega \cdot \beta_p)$.
6. The numerical values of coefficients ($v$ and $\omega$) are calculated during the construction of $H_C$ and $H_M$, in the classical domain.
7. The numerical values of $\gamma$ and $\beta$ are initially randomized as a set $[\gamma_1 \ldots \gamma_p, \beta_1 \ldots \beta_p]$ for $H_C$ and $H_M$, respectively, in the classical domain.
8. The quantum circuit of QAOA (H$\{H_C H_M\}^p$) is executed with a quantum processing unit (QPU), and then measured (in the classical domain) for solutions depending on the chosen values of $\gamma$ and $\beta$.
9. The measured solutions (as the *energy cost* of QAOA [1, 2]), the chosen values of $\gamma$ and $\beta$ (as the *parameters* of QAOA), and the Hamiltonians ($H_C$ and $H_M$ as an *objective function*) are fed to a classical optimization minimizer [9-11]. This minimizer re-calculates the numerical values of these parameters based on the energy cost from the objective function, and updates $H_C$ and $H_M$ of the quantum circuit of QAOA with a new set of optimized numerical values of $\gamma$ and $\beta$, respectively.
10. For a number of objective function evaluations, steps 8 and 9 are repeated concurrently between a QPU and a minimizer, until finding all optimized approximated solutions for a MaxCut problem.

The aforementioned ten steps tell why QAOA is considered a variational hybrid classical-quantum algorithm [12-14], and the wording "variational" is equivalent to the meaning of "ansatz".

Different research and proposed implementations of QAOA are fundamentally concentrated on: (i) solving graphs, $k$-SAT, and MaxCut problems [3, 4, 15, 16], where $k \geq 3$ inputs, (ii) finding the optimized numerical values of $\gamma$ and $\beta$ for a fewer number of optimization steps and QAOA repetitions ($p$) [17-21], and (iii) developing variants of $H_M$ for a better QAOA in finding all solutions [22-27].

In contrast, in this paper, we propose a new methodology to solve classical Boolean problems as Hamiltonians ($H_C$ and $H_M$) using QAOA. For that, this proposed methodology is termed the "Boolean-Hamiltonians Transform for QAOA (BHT-QAOA)", which can be summarized as follows. Firstly, an arbitrary classical Boolean problem is constructed as a quantum Boolean oracle [8, 28]. This constructed oracle is built in arbitrary structures, such as Product-Of-Sums (POS) [29, 30], Sum-Of-Products (SOP) [29, 31],

Exclusive-or Sum-Of-Products (ESOP) [32, 33], XOR-Satisfiability (CNF-XOR SAT and DNF-XOR SAT) [34, 35], Algebraic Normal Form (ANF) (or Reed-Muller expansion) [36, 37], just to name a few. Secondly, this quantum Boolean oracle (in any structure) is converted into its equivalent quantum Boolean oracle in ESOP structure, unless it was initially constructed in ESOP structure. Thirdly, this quantum Boolean oracle in ESOP structure is transformed into its equivalent quantum Phase oracle [8, 28]. Fourthly, the Hamiltonians ($H_C$ and $H_M$) of QAOA are generated from this transformed quantum Phase oracle, based on the modified composition rules originally presented by Hadfield [38]. Finally, all the above-mentioned ten steps of QAOA are performed in sequence using our generated Hamiltonians ($H_C$ and $H_M$).

A quantum Boolean oracle is easier and more straightforward in designing an arbitrary classical Boolean problem than a quantum Phase oracle, because (i) the quantum Boolean-based gates can be directly realized using the truth tables or De Morgan's Laws [29] from their equivalent classical Boolean gates, and (ii) the quantum Boolean-based gates and their qubits can be easily analyzed using classical Boolean logic, i.e., a Boolean logic of '0' represents a quantum state of |0⟩ and a Boolean logic of '1' represents a quantum state of |1⟩. For ease of description, the quantum Boolean oracle and the quantum Phase oracle will be simply denoted as the "Boolean oracle" and the "Phase oracle", respectively.

In our BHT-QAOA, converting a Boolean oracle (in any structure) into a Phase oracle will: (i) remove all ancilla qubits (including the output qubit), i.e., the total number of utilized qubits will be dramatically reduced to the number of input qubits only, and (ii) omit the mirror (as the uncomputing part) of a Boolean oracle, i.e., the total number of quantum gates are minimized for the quantum circuit of Phase oracle, depending on the initial construction of a Boolean oracle for a classical Boolean problem.

In this paper, arbitrary classical Boolean problems (as applications) are converted into Boolean oracles in various structures, and these Boolean oracles are then solved using BHT-QAOA for $p$ repetitions, with an IBM QPU and SciPy optimization minimizer [39]. These applications are: (i) arbitrary classical problem in POS structure, (ii) arbitrary classical problem in SOP structure, (iii) arbitrary classical problem in ESOP structure, (iv) 2×2 Sudoku game, and (v) 4-bit conditioned half-adder digital circuit. Eventually, BHT-QAOA successfully finds all optimized approximated solutions for these applications, as a proof of concept for utilizing BHT-QAOA to solve arbitrary classical Boolean problems in the hybrid classical-quantum domain.

## 2  Methodology

Let us start with an illustrative example. An arbitrary classical Boolean problem is given, as stated in Eq. (1) below, and then designed as a Boolean oracle in POS structure, as illustrated in Figure 1(a). Subsequently, the following subsections discuss how to: (i) convert this Boolean oracle in POS structure into the Boolean oracle in ESOP structure, (ii) transform the Boolean oracle in ESOP structure into the Phase oracle, (iii) generate the Hamiltonians ($H_C$ and $H_M$) based on the transformed Phase oracle, and (iv) construct the overall architecture of BHT-QAOA for $H_C$ and $H_M$ in $p$ repetitions, where $p ≥ 1$. Note that, for other classical Boolean problems, differently designed Boolean oracles (in any structure) can follow the same steps of conversion, transformation, and generation stated in these subsections.

$$(a \lor b \lor \neg c) \land (\neg a \lor c) \land (\neg b \lor c) \tag{1}$$

## 2.1 Converting Boolean oracles from any structure to ESOP structure

There are many synthesis methods to convert a Boolean oracle (in any structure) to its equivalent Boolean oracle in ESOP structure. These different synthesis methods include ESOP synthesis [32], Karnaugh map synthesis [29], binary decision diagram (BDD) synthesis [40, 41], just to name a few. From Eq. (1) above, because this Boolean oracle in POS structure is simple, the Karnaugh map synthesis is utilized to convert it to the Boolean oracle in ESOP structure, as stated in Eq. (2) below and illustrated in Figure 1(c)(d), and the final quantum circuit of this Boolean oracle in ESOP structure is shown in Figure 1(b). Note that (i) there is no need to optimize this Boolean oracle in ESOP structure, e.g., removing the identical neighbored quantum gates, since all these gates are required for generating Hamiltonians ($H_C$ and $H_M$) and calculating their coefficients ($v$ and $\omega$), respectively, and (ii) the *fqubit* is the output ancilla qubit (also called the functional qubit) of a Boolean oracle (in any structure).

$$(\neg a \wedge \neg b \wedge \neg c) \oplus (a \wedge \neg b \wedge c) \oplus (b \wedge c) \quad (2)$$

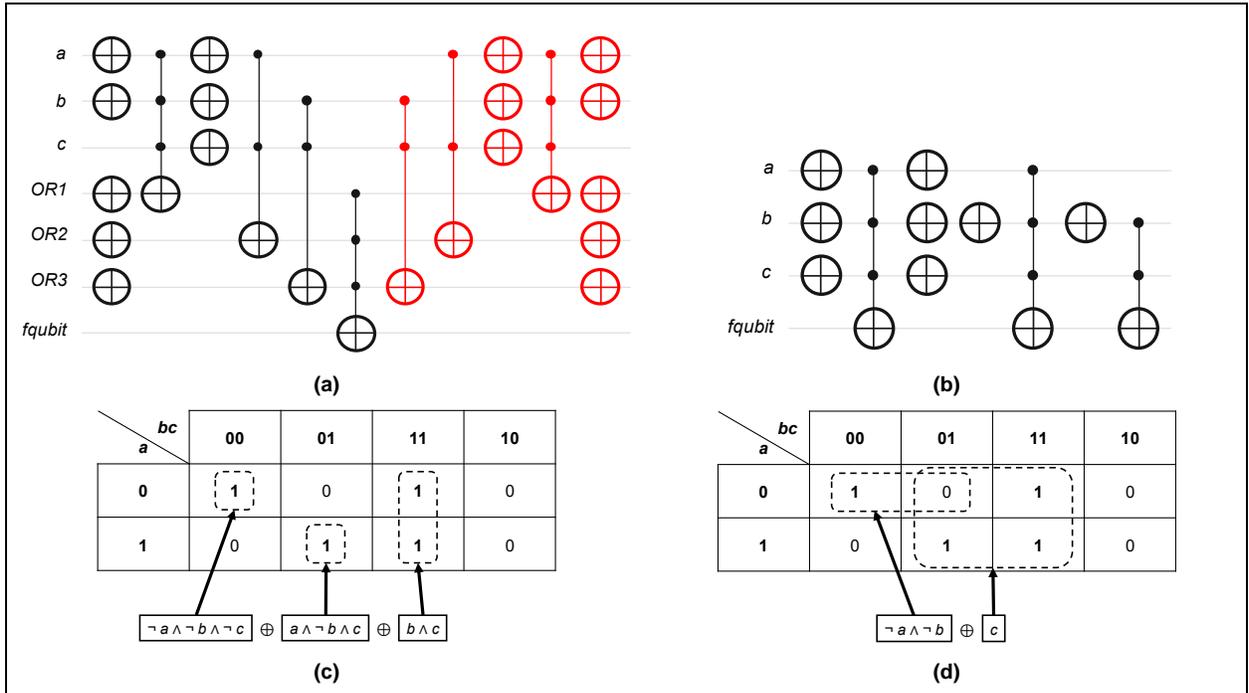

**Figure 1.** The quantum circuits and Karnaugh map syntheses for the classical Boolean problem $(a \vee b \vee \neg c) \wedge (\neg a \vee c) \wedge (\neg b \vee c)$: **(a)** the Boolean oracle in POS structure, where *OR*1 represents the term $(a \vee b \vee \neg c)$, *OR*2 represents the term $(\neg a \vee c)$, *OR*3 represents the term $(\neg b \vee c)$, *fqubit* performs all AND operations ($\wedge$), and the quantum gates in red denote the mirror (as the uncomputing part) of this Boolean oracle that resets all ancilla qubits to their initial |0⟩ states, **(b)** the Boolean oracle in DSOP (as a special case of ESOP) structure of $(\neg a \wedge \neg b \wedge \neg c) \oplus (a \wedge \neg b \wedge c) \oplus (b \wedge c)$, where the *fqubit* performs all XORing operations ($\oplus$), **(c)** the Karnaugh map synthesis from the Boolean oracle in (a) to the Boolean oracle in (b), based on only grouping all true minterms '1' (as solutions), and **(d)** the Karnaugh map synthesis from the Boolean oracle in (a) to its equivalent Boolean oracle in ESOP structure $(\neg a \wedge \neg b) \oplus (c)$, based on grouping all true minterms '1' (as solutions) with one false minterm '0' (as an XORed solution). Note that (i) all ancilla qubits (including *fqubit*) are initially set to the |0⟩ states, (ii) the DSOP structure is a non-minimized ESOP structure and a structure minimization method assumes to reduce the number of $n$-bit Toffoli gates, where $n \geq 3$ qubits, (iii) all mirrored gates and ancilla qubits (except for *fqubit*) are removed when a Boolean oracle is in ESOP (or DSOP) structure, and (iv) there is no need to remove neighbored Pauli-X (X) gates, since all these gates are required for generating the Hamiltonians ($H_C$ and $H_M$) and calculating their coefficients ($v$ and $\omega$), respectively.

In general, the Karnaugh map synthesis for a Boolean oracle (from any structure to ESOP structure) can be summarized in the following steps. Note that the following steps may not create the minimized ESOP structure, as shown in Figure 1(d), since these steps usually create the DSOP (Disjoint Sum-Of-Products) structure, as shown in Figure 1(c), which is an expensive structure as compared to the minimized ESOP structure based on the numbers of $n$-bit Toffoli gates, where $n \geq 3$ qubits. However, in this subsection, we just want to simply illustrate how to convert a Boolean oracle (in any structure) to its equivalent Boolean oracle in ESOP (or DSOP) structure.

1. Sketch an empty Karnaugh map for the literals, e.g., *a*, *b*, and *c*, with their binary Gray codes.
2. Evaluate a Boolean oracle (in any structure), as solutions (for the true minterms of '1') and non-solutions (for the false minterms of '0').
3. Group all solutions together from step 2, as 1-cell groups, 2-cell groups, 4-cell groups, etc.
4. Formulate each group from step 3, as products (∧) of literals.
5. XOR (⊕) all formulated groups together from step 4, as one ESOP (or DSOP) structure.

## 2.2 Transforming Boolean oracles in ESOP structure to Phase oracles

In this paper, we utilize the technique originally discussed by Figgatt *et al.* [28] for transforming 4-bit Toffoli (CCCX) gates to 3-bit controlled-Z (CCZ) gates, for Grover's algorithm of single-solution [5-8]. We efficiently generalize their technique to involve the Feynman (CX) and $n$-bit Toffoli gates, where $n \geq 3$ qubits, for transforming Boolean oracles in ESOP structure to their equivalent Phase oracles, as presented in the following three rules and demonstrated in Figure 2. For that, we termed these rules the "generalized transformation rules".

**Rule 1:** A Feynman (CX) gate is transformed into a Pauli-Z (Z) gate, when Eq. (3) stated below is satisfied as demonstrated in Figure 2(a), where *j* is the index of an input qubit (q) and *fqubit* initially sets to the |0⟩ state. Note that the left-side of Eq. (3) is the output of a CX gate, and its right-side is the negated output of a Z gate.

$$q_j \oplus \textit{fqubit} = -(-1)^{q_j} \tag{3}$$

**Rule 2:** A Toffoli gate is transformed into a controlled-Z (CZ) gate, when Eq. (4) stated below is satisfied as demonstrated in Figure 2(b), where *j* and *k* are the indices of input qubits (q), and *fqubit* initially sets to the |0⟩ state. Note that the left-side of Eq. (4) is the output of a Toffoli gate, and its right-side is the negated output of a CZ gate.

$$\left(q_j \wedge q_k\right) \oplus \textit{fqubit} = -(-1)^{q_j \cdot q_k} \tag{4}$$

**Rule 3:** An *n*-bit Toffoli gate is transformed into an (*n*–1)-bit multi-controlled Z (MCZ) gate, when Eq. (5) stated below is satisfied as demonstrated in Figure 2(c), where *j* is the index of an input qubit (q), *fqubit* initially sets to the |0⟩ state, and *n* ≥ 3 qubits (q + *fqubit*). Note that the left-side of Eq. (5) is the output of an *n*-bit Toffoli gate, and its right-side is the negated output of an (*n*–1)-bit MCZ gate.

$$\left(\bigwedge_{j=1}^{n-1} q_j\right) \oplus fqubit = -(-1)^{\prod_{j=1}^{n-1} q_j} \tag{5}$$

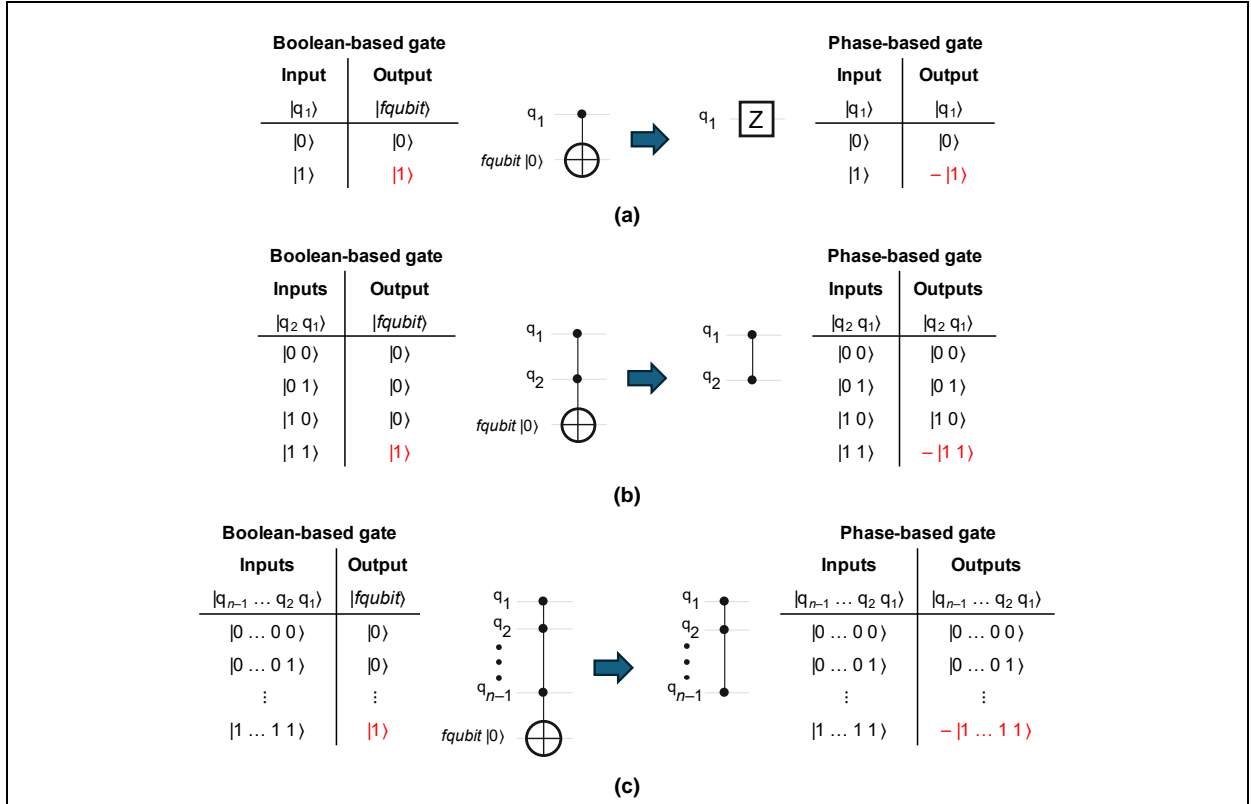

**Figure 2.** Schematics of our generalized transformation rules from the quantum Boolean-based gates (Feynman, Toffoli, and *n*-bit Toffoli gates) to the quantum Phase-based gates (Z, CZ, and (*n*–1)-bit MCZ gates) with their truth tables as proofs, where Z is the Pauli-Z gate, CZ is the controlled-Z gate, MCZ is the multi-controlled Z gate, *n* ≥ 3 qubits (q inputs + *fqubit*), and texts in red indicate the solutions: **(a)** Rule 1 transforms a Feynman (CX) gate into a Z gate, **(b)** Rule 2 transforms a Toffoli gate into a CZ gate, and **(c)** Rule 3 transforms an *n*-bit Toffoli gate to an (*n*–1)-bit MCZ gate. Note that the total number of qubits is reduced by one when transforming a quantum Boolean-based gate into a quantum Phase-based gate, i.e., the *fqubit* is removed.

After applying our generalized transformation rules on the Boolean oracle in ESOP (or DSOP) structure (as stated in Eq. (2) above and shown in Figure 1(b)), the resultant quantum circuit of Phase oracle is then simply constructed, as shown in Figure 3(a). Additionally, Table 1 summarizes the advantages of converting the Boolean oracle in POS structure to the Boolean oracle in ESOP structure, and transforming the Boolean oracle in ESOP structure to the Phase oracle, in the context of: (i) the utmost removal of all ancilla qubits (including the *fqubit*), i.e., the width of the final quantum circuit is reduced, and (ii) the dramatic minimization of multi-controlled quantum gates, i.e., the depth of the final quantum circuit is shrunk.

**Table 1.** Comparison of the number of qubits and quantum gates for the Boolean and Phase oracles for the classical Boolean problem of $(a \lor b \lor \neg c) \land (\neg a \lor c) \land (\neg b \lor c)$, after applying the generalized transformation rules.

| Oracular problem | Number of qubits | | | Number of multi-controlled gates | | | Quantum circuit required a mirror? |
|---|---|---|---|---|---|---|---|
| | Inputs | Ancillae (incl. *fqubit*) | Total | Feynman (CX) | 3-bit Toffoli | 4-bit Toffoli | |
| Boolean oracle (POS structure) | 3 | 4 | 7 | 0 | 4 | 3 | Yes |
| Boolean oracle (ESOP structure) | 3 | 1 | 4 | 0 | 1 | 2 | No |
| Phase oracle | 3 | 0 | 3 | 1 (as CZ) | 2 (as CCZ) | 0 | No |

### 2.3 Generating Hamiltonians ($H_C$ and $H_M$) from Phase oracles

Hadfield discussed, in [38], the composition rules for generating $H_C$ from a set of Hamiltonians ($H_f$), which represent a variety of Boolean functions as simpler clauses, as stated in Table 2. In this paper, to generate the Hamiltonians ($H_C$ and $H_M$) from Phase oracles, we generalize some of Hadfield's Boolean-based composition rules ($H_f$) to Phase-based composition rules, which we termed the "generalized composition rules ($H_g$)". Based on the three generalized transformation rules stated above, four generalized composition rules ($H_g$) are derived from $H_f$, as expressed in Table 3, where Rules 1, 2, and 3 of $H_g$ are simply inverting the signs (±) of their corresponding $H_f$ (for both identity 'I' and RZ gates).

**Rule 1:** Performs Rule 1 of the generalized transformation rules, as stated in Eq. (3) above and previously shown in Figure 2(a).

**Rule 2:** Performs Rule 2 of the generalized transformation rules, as stated in Eq. (4) above and previously shown in Figure 2(b).

**Rule 3:** Performs Rule 3 of the generalized transformation rules, as stated in Eq. (5) above and previously shown in Figure 2(c).

**Rule 4:** Inverts all signs (±) of generated RZ gates in $H_g$ (Rules 1, 2, and 3), since the Pauli-X (X) gates are proposed here to invert the phases of qubits in a Hamiltonian ($H_g$), and not their states.

**Table 2.** Some of Hadfield's Boolean-based composition rules ($H_f$) [38], where I is the identity gate, $n$ is the total number of all qubits (q inputs + *fqubit*), $j$ and $k$ are the indices of q inputs, $Z_j$ is the RZ gate applied on $q_j$, and *fqubit* initially sets to the $|0\rangle$ state.

| Gate | Type | f(x) | $H_f$ |
|---|---|---|---|
| Feynman (CX) | Boolean | $q_j \oplus fqubit = q_j$ | $\frac{1}{2} I - \frac{1}{2} Z_j$ |
| 3-bit Toffoli (CCX) | Boolean | $q_j \land q_k$ | $\frac{1}{4} I - \frac{1}{4}(Z_j + Z_k - Z_j Z_k)$ |
| $n$-bit Toffoli ($C^{n-1}X$) | Boolean | $\bigwedge_{j=1}^{n-1} q_j$ | $\frac{1}{2^{n-1}} \prod_{j=1}^{n-1} (I - Z_j)$ |

**Table 3.** Our proposed generalized composition rules ($H_g$) for Phase oracles, where I is the identity gate, $n$ is the total number of input qubits (q), $j$ and $k$ are the indices of q input qubits, $Z_j$ is the RZ gate applied on $q_j$, $Q = \{q_j, q_k \ldots q_j q_k \ldots\}$, and $Z_Q = \{Z_j, Z_k \ldots Z_j Z_k \ldots\}$.

| | Gate | Type | g(x) | $H_g$ |
|---|---|---|---|---|
| **Rule 1:** | Pauli-Z (Z) | Phase | $(-1)^{q_j}$ | $-\frac{1}{2}I + \frac{1}{2}Z_j$ |
| **Rule 2:** | Controlled-Z (CZ) | Phase | $(-1)^{q_j \cdot q_k}$ | $-\frac{1}{4}I + \frac{1}{4}(Z_j + Z_k - Z_j Z_k)$ |
| **Rule 3:** | $n$-controlled Z ($C^n Z$ or MCZ) | Phase | $(-1)^{\prod_{j=1}^{n} q_j}$ | $\frac{1}{2^n} \prod_{j=1}^{n} (-1)^{j+1}(I - Z_j)$ |
| **Rule 4:** | Pauli-X (X) | Phase | $(-1)^{\forall j \in Q}$ | Invert signs (±) of all $j$th qubit in $Z_Q$ |

Consequently, from Table 3, the four generalized compositions rules ($H_g$) will be then directly applied to the Phase oracle (shown in Figure 3(a)) to generate $H_C$ and calculate its $v$ coefficient, as demonstrated in Figure 3(b) and expressed in the following steps (starting from the left-side of this figure).

1. Construct one $H_g$ for one Z, CZ, or MCZ gate, using Rule 1, Rule 2, or Rule 3, respectively.
2. If there are Pauli-X (X) gates (with their mirrored gates) surrounding that Z, CZ, or MCZ gate in step 1, then apply Rule 4 on $H_g$ from step 1 to construct a new $H_g$.
3. Repeat steps 1 and 2 for another $H_g$, until there are no remaining Z, CZ, or MCZ gates.
4. Concatenate all constructed $H_g$ into one Hamiltonian, which is $H_C$.
5. Calculate (add or subtract) all the identical terms of $H_C$ to find the $v$ coefficient.

From Table 3 and Figure 3(b), the constructed Hamiltonians ($H_g$) are presented in Eq. (6) below. Since $H_C = H_{g2} + H_{g4} + H_{g5}$, then $H_C$ is simply constructed as stated in Eq. (7) below.

$$H_{g1} = \frac{1}{8}((I-Z_a)(-I+Z_b)(I-Z_c)) = \frac{1}{8}(-I+Z_a+Z_b+Z_c-Z_aZ_b-Z_aZ_c-Z_bZ_c+Z_aZ_bZ_c)$$

$$H_{g2} = \frac{1}{8}(-I-Z_a-Z_b-Z_c-Z_aZ_b-Z_aZ_c-Z_bZ_c-Z_aZ_bZ_c)$$

$$H_{g3} = \frac{1}{8}((I-Z_a)(-I+Z_b)(I-Z_c)) = \frac{1}{8}(-I+Z_a+Z_b+Z_c-Z_aZ_b-Z_aZ_c-Z_bZ_c+Z_aZ_bZ_c)$$

$$H_{g4} = \frac{1}{8}(-I+Z_a-Z_b+Z_c+Z_aZ_b-Z_aZ_c+Z_bZ_c-Z_aZ_bZ_c)$$

$$H_{g5} = \frac{1}{4}((I-Z_b)(-I+Z_c)) = \frac{1}{4}(-I+Z_b+Z_c-Z_bZ_c) \tag{6}$$

$$H_C = -\frac{1}{2}I + \frac{1}{4}Z_c - \frac{1}{4}(Z_aZ_c + Z_bZ_c) - \frac{1}{4}Z_aZ_bZ_c \tag{7}$$

Hence, from Eq. (7) above, $v = [v_1, v_2, v_3, v_4] = [-\frac{1}{2}, \frac{1}{4}, -\frac{1}{4}, -\frac{1}{4}]$, where $v_1$ is for all non-connected input qubits (as the identity 'I'), $v_2$ is for $c$ input qubit only, i.e., $RZ_c(0.25 \cdot \gamma)$, $v_3$ is for only two connected input qubits $\{a$ and $c$; $b$ and $c\}$, i.e., $RZ_aZ_c(-0.25 \cdot \gamma)$ and $RZ_bZ_c(-0.25 \cdot \gamma)$, and $v_4$ is for all connected input qubits $\{a, b, c\}$, i.e., $RZ_aZ_bZ_c(-0.25 \cdot \gamma)$. For a practical quantum implementation of QAOA as well as our proposed BHT-QAOA, $H_C$ is re-written to the following format for the three qubits $\{a, b, c\}$, as stated in Eq. (8) below, where 'ZII' is equivalent to $RZ_c$, 'ZIZ' is equivalent to $RZ_aZ_c$, 'ZZZ' is equivalent to $RZ_aZ_bZ_c$, and so on.

$$H_C = (['III', 'ZII', 'ZIZ', 'ZZI', 'ZZZ'], \ coeffs = [-0.25, 0.25, -0.25, -0.25, -0.25]) \tag{8}$$

Because $β$ (as a set of rotational angles of $H_M$) restrictively rotates between [0, π] [1, 2], we set $ω$ to cover all the range between [0, 2π], to find all optimized approximated solutions for an arbitrary classical Boolean problem. In other words, $ω$ (as the coefficient of $β$) is initially set to '2.0' for all $n$ numbers of RX($ω·β$) gates in $H_M$, where $n$ is the total number of input qubits. Similar to the practical implementation of $H_C$, $H_M$ is re-written to the following format for the three qubits {$a$, $b$, $c$}, as stated in Eq. (9) below, where 'XII' is equivalent to RX$_c$, 'IXI' is equivalent to RX$_b$, and 'IIX' is equivalent to RX$_a$.

$$H_M = (['XII', 'IXI', 'IIX'], \ coeffs = [2.0, 2.0, 2.0]) \tag{9}$$

From Eq. (8) and Eq. (9) above, the quantum circuit of BHT-QAOA for the Hamiltonians ($H_C$ and $H_M$) of the classical Boolean problem ($a ∨ b ∨ ¬ c$) ∧ ($¬ a ∨ c$) ∧ ($¬ b ∨ c$) is eventually constructed for one repetition ($p$ = 1), as illustrated in Figure 3(c).

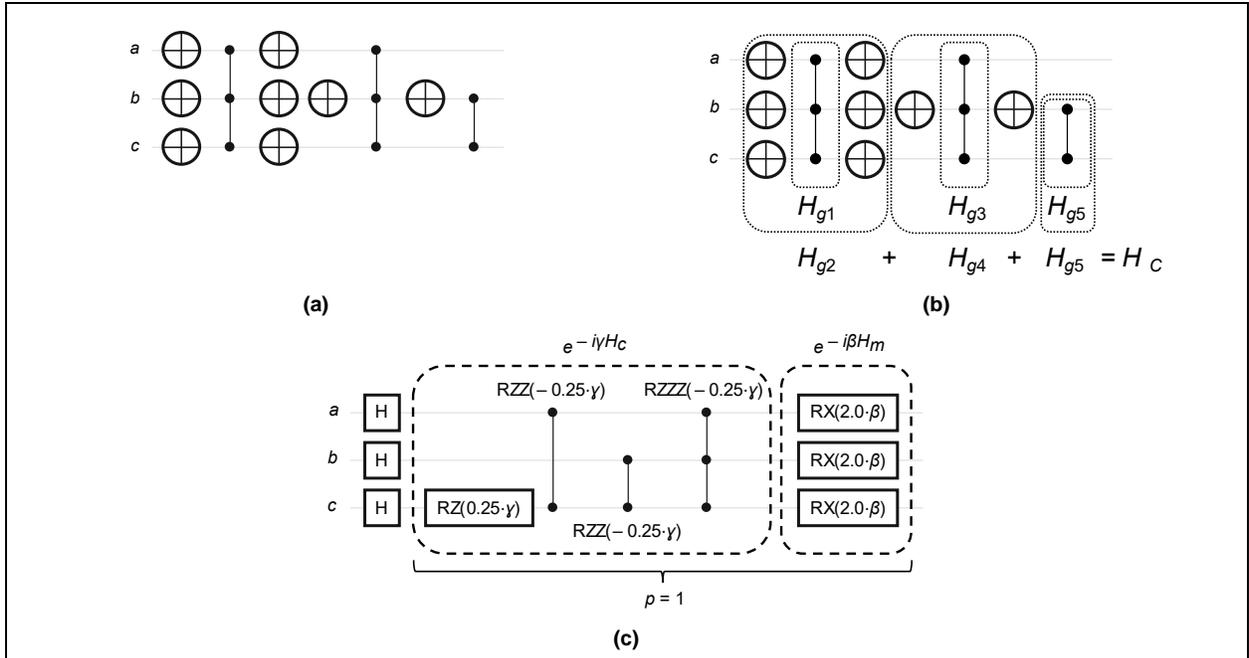

**Figure 3.** Quantum circuits for the classical Boolean problem of ($a ∨ b ∨ ¬ c$) ∧ ($¬ a ∨ c$) ∧ ($¬ b ∨ c$) as: **(a)** the Phase oracle after applying our generalized transformation rules on the Boolean oracle in ESOP structure ($¬ a ∧ ¬ b ∧ ¬ c$) ⊕ ($a ∧ ¬ b ∧ c$) ⊕ ($b ∧ c$), such that (i) two 4-bit Toffoli gates are transformed into two 3-bit MCZ gates, (ii) one 3-bit Toffoli gate is transformed into one 2-bit CZ gate, (iii) all ancilla qubits (including *fqubit*) are removed, and (iv) there is no mirror (as the uncomputing part) of this Phase oracle, **(b)** the sequences of Hamiltonians ($H_g$) using our generalized composition rules, to directly generate $H_C$ and calculate its $v$ coefficient from the Phase oracle in (a), where $H_C$ = ($H_{g1}$ ➔ $H_{g2}$) + ($H_{g3}$ ➔ $H_{g4}$) + $H_{g5}$, and **(c)** the BHT-QAOA of two Hamiltonians ($H_C$ and $H_M$) in one repetition ($p$), where H is the Hadamard gate, $H_C$ = (['III', 'ZII', 'ZIZ', 'ZZI', 'ZZZ'], $coeffs$ = [− 0.25, 0.25, − 0.25, − 0.25, − 0.25]), and $H_M$ = (['XII', 'IXI', 'IIX'], $coeffs$ = [2.0, 2.0, 2.0]). Note that CZ is the controlled Z gate and MCZ is the multi-controlled Z gate.

## 2.4 Architecture of BHT-QAOA

After transforming an arbitrary classical Boolean problem to the Hamiltonians ($H_C$ and $H_M$) and calculating their coefficients ($v$ and $\omega$), respectively, the numerical values of $\gamma$ and $\beta$ are initially randomized and then plugged into the quantum circuit of BHT-QAOA for first execution, as shown in Figure 4. Subsequently, the SciPy optimization minimizer [39] optimizes these numerical values of $\gamma$ and $\beta$ for better-approximated solutions, in a number of function evaluations (*nfev*), by utilizing three parameters as follows.

1. Hamiltonians ($H_C$ and $H_M$ in a number of *p*), as the "objective function" needs to be minimized.
2. Previously measured solutions of BHT-QAOA, as the "energy cost" of the objective function.
3. Previously calculated $\gamma$ and $\beta$, as their "numerical values" need to be optimized.

For the SciPy optimization minimizer, we use the constrained optimization by linear approximation (COBYLA) algorithm [9, 11, 39], which is a parametric iterative method for derivative-free constrained optimization that updates and minimizes the linear approximation values (as $\gamma$ and $\beta$) for an objective function (as Hamiltonians of lower energy cost). On the other hand, our future work will focus on finding a cost-effective minimizer algorithm that efficiently implements BHT-QAOA with a smaller value of *p* (in the quantum domain) and fewer *nfev* (in the classical domain).

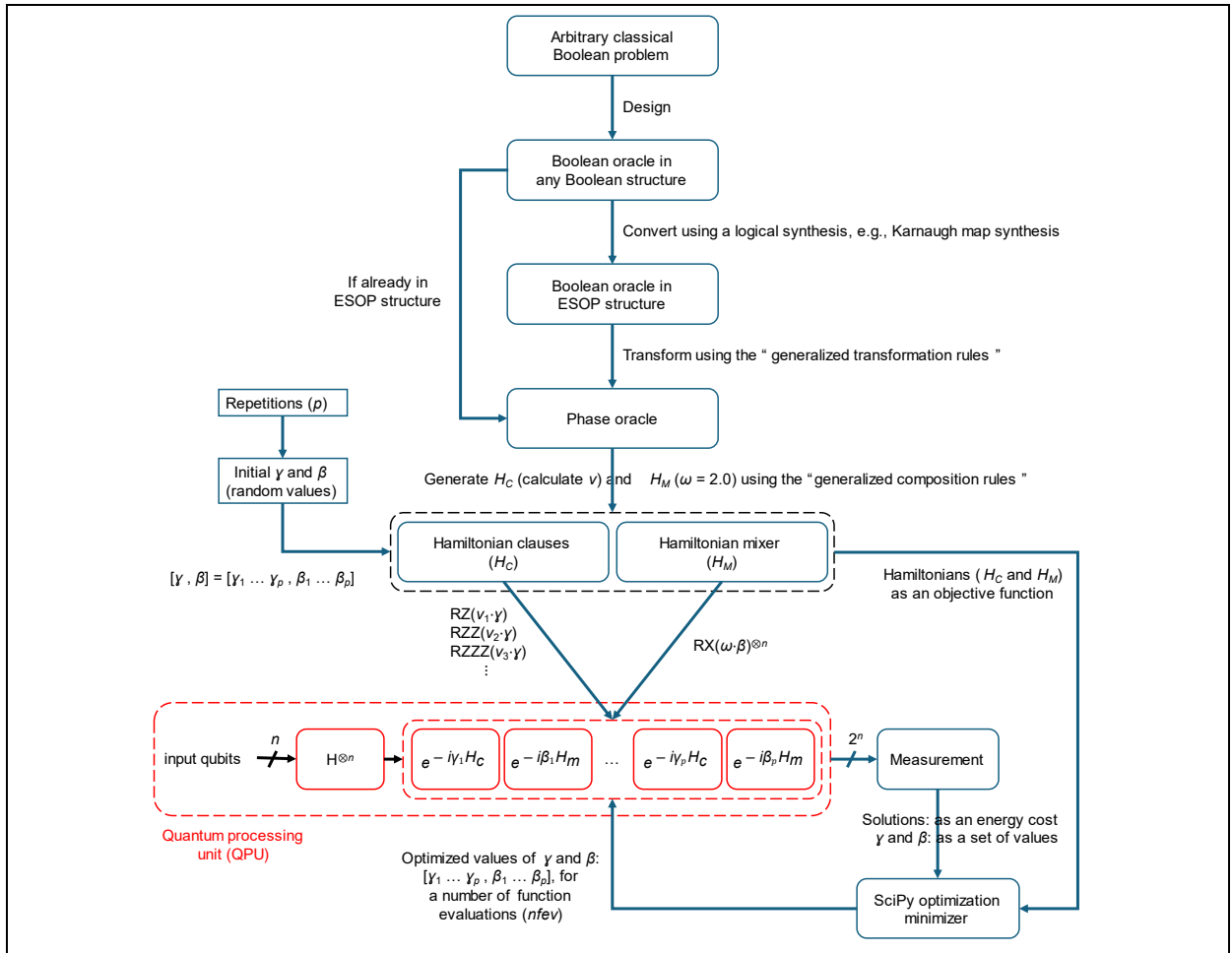

**Figure 4.** The architecture of our proposed Boolean-Hamiltonians Transform for QAOA (BHT-QAOA) to solve arbitrary classical Boolean problems as Hamiltonians ($H_C$ and $H_M$). The BHT-QAOA is mainly grouped into two processing domains: (i) the classical processing domain as denoted by blue, and (ii) the quantum processing domain as denoted by red.

## 3 Results

Arbitrary Boolean problems (as applications) are designed as Boolean oracles (in different structures), and these Boolean oracles are then solved using our proposed BHT-QAOA for *p* repetitions, where *p* ≥ 1. In our experiments, the `ibm_brisbane` [42] QPU of 127 qubits was utilized to perform the quantum processing domain of BHT-QAOA, i.e., to execute the quantum circuit of an application in *p* repetitions, while the SciPy optimization minimizer was used to perform the classical processing domain of BHT-QAOA, i.e., to optimize the numerical values of *γ* and *β* based on the minimized energy cost of their Hamiltonians ($H_C$ and $H_M$) as the objective function, for a number of function evaluations (*nfev*).

Because of the access limitations of our IBM Quantum Platform account, the architecture of BHT-QAOA (as previously illustrated in Figure 4) was completely simulated in the classical domain using IBM quantum libraries: `Qiskit`, `AerSimulator`, and `Aer-EstimatorV2` [42-44], for 1024 resampling times, which is the so-called "shots" [45]. After this classically simulated BHT-QAOA, the final optimized numerical values of *γ* and *β* (from SciPy optimization minimizer) were plugged into their respective Hamiltonians ($H_C$ and $H_M$) of the quantum circuit for an application, in which it is then executed once with `ibm_brisbane` QPU.

Due to the limited physical connectivity of four neighboring qubits for the recent layouts of IBM QPUs, the designed Boolean oracles (in various structures) for the following applications must have *n* input qubits and *m* ancilla qubits (including *fqubit*), where 2 ≤ *n* ≤ 4 and *m* ≥ 1. Note that the *m* ancilla qubits: (i) do not affect the fidelity of the final optimized approximated solutions of applications with BHT-QAOA, and (ii) have no relation to the limited physical connectivity of any IBM QPU, because all these designed Boolean oracles are transformed into Phase oracles and then to Hamiltonians ($H_C$ and $H_M$), which do not have any ancilla qubits in their quantum circuits, i.e., all *m* ancilla qubits are removed. Figure 5 demonstrates the classical representations and the quantum circuits of Boolean oracles for these applications as follows.

1. Arbitrary Boolean problem in POS structure, as stated in Eq. (1) above and shown in Figure 1(a).
2. Arbitrary Boolean problem in SOP structure, as stated in Eq. (10) below and shown in Figure 5(a).
3. Arbitrary Boolean problem in ESOP structure, as stated in Eq. (11) below and shown in Figure 5(b).
4. 2×2 Sudoku game, which is the constraints satisfaction problem – satisfiability (CSP-SAT) [46, 47], as stated in Eq. (12) below and shown in Figure 5(c)(d).
5. 4-bit conditioned half-adder digital circuit, which is ORing two 1-bit sums and then ANDing them with one 1-bit carry-out, as stated in Eq. (13) below and shown in Figure 5(e)(f).

$$(a \wedge b \wedge \neg c) \vee (\neg a \wedge c) \vee (\neg b \wedge c) \qquad (10)$$

$$(a \wedge b \wedge \neg c) \oplus (\neg a \wedge c) \oplus (\neg b \wedge c) \qquad (11)$$

$$(cell_1 \oplus cell_2) \wedge (cell_1 \oplus cell_3) \wedge (cell_2 \oplus cell_4) \wedge (cell_3 \oplus cell_4) \qquad (12)$$

$$[(a_0 \oplus b_0) \vee ((a_0 \wedge b_0) \oplus (a_1 \oplus b_1))] \wedge [(a_1 \wedge b_1) \vee ((a_0 \wedge b_0) \wedge (a_1 \oplus b_1))] \qquad (13)$$

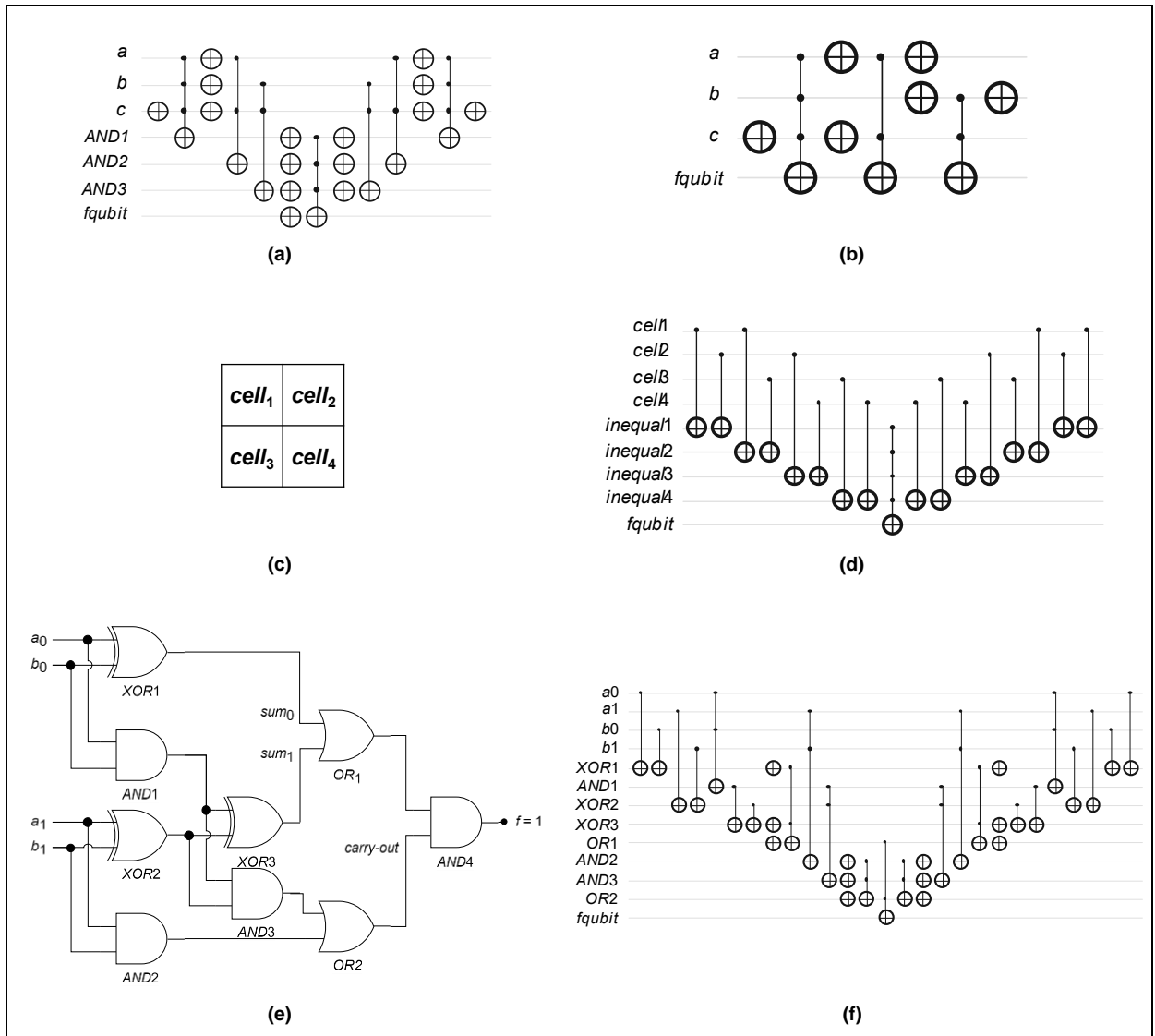

**Figure 5.** The classical representations and the Boolean oracles for arbitrary Boolean problems: **(a)** the Boolean oracle in SOP structure of Eq. (10) above, **(b)** the Boolean oracle in ESOP structure of Eq. (11) above, **(c)** the board layout of 2×2 Sudoku game, **(d)** the Boolean oracle in CNF-XOR SAT structure of Eq. (12) above, **(e)** the digital circuit of 4-bit conditioned half-adder for two 2-bit numbers (A = $a_1a_0$ and B = $b_1b_0$), and **(f)** the Boolean oracle in mixed structures of Eq. (13) above.

Table 4 states the removal of all ancilla qubits (including *fqubit*) and the reduction number of quantum gates, after applying our proposed generalized transformation rules to transform Boolean oracles (in any structure) into cost-effective Phase oracles. These Phase oracles are then transformed into the Hamiltonians ($H_C$ and $H_M$) of BHT-QAOA. In Table 4, (i) a Pauli-X (X) gate in a Boolean oracle remains as-is for a Phase oracle, (ii) a Feynman (CX) gate in a Boolean oracle is equivalent to the controlled Pauli-Z (CZ) gate in a Phase oracle, and (iii) an *n*-bit Toffoli gate in a Boolean oracle is equivalent to an *n*-bit MCZ gate in a Phase oracle, where $n \geq 3$ qubits.

**Table 4.** The effect of generalized transformation rules on removing all ancilla qubits (including *fqubit*) and reducing the numbers of quantum gates, when Boolean oracles (in any structure) are transformed into Phase oracles for BHT-QAOA.

| Qubits and quantum gates | Classical Boolean problems (applications) {entities in Boolean oracles → entities in Phase oracles} | | | | |
|---|---|---|---|---|---|
| | Arbitrary problem in POS | Arbitrary problem in SOP | Arbitrary problem in ESOP | 2×2 Sudoku game | 4-bit conditioned half-adder circuit |
| **Input qubits** | 3 → 3 | 3 → 3 | 3 → 3 | 4 → 4 | 4 → 4 |
| **Ancilla qubits (incl. *fqubit*)** | 4 → 0 | 4 → 0 | 1 → 0 | 5 → 0 | 9 → 0 |
| **Pauli-X (X)** | 16 → 8 | 15 → 6 | 6 → 6 | 0 → 8 | 12 → 2 |
| **Feynman (CX)** | 0 → 1 | 0 → 1 | 0 → 2 | 16 → 0 | 12 → 0 |
| **3-bit Toffoli** | 4 → 2 | 4 → 2 | 2 → 1 | – | 11 → 1 |
| **4-bit Toffoli** | 3 → 0 | 3 → 0 | 1 → 0 | 0 → 2 | 0 → 1 |
| **5-bit Toffoli** | – | – | – | 1 → 0 | – |

On the one hand, the classically simulated BHT-QAOA successfully optimizes the numerical values of $γ$ and $β$ and finds all approximated solutions for these applications, as a proof of concept for utilizing our proposed BHT-QAOA in solving arbitrary classical Boolean problems in the simulated classical-quantum domain. On the other hand, the quantum circuit of every application (using the simulated optimized numerical values of $γ$ and $β$) is executed once with `ibm_brisbane` QPU for solutions fidelity, as a proof of concept for utilizing BHT-QAOA in solving arbitrary classical Boolean problems in the hybrid classical-quantum domain as well. Figure 6 depicts the final measured solutions for every application from the real quantum executions using `ibm_brisbane` QPU.

Note that the required *nfev* for the SciPy optimization minimizer varies and fluctuates, since such a minimizer mainly depends on: (i) the initially randomized numerical values of $γ$ and $β$ for $H_C$ and $H_M$, respectively, and (ii) the noisy simulation models (`AerSimulator` and `Aer-EstimatorV2`) that are utilized for simulating BHT-QAOA, in the classical domain.

Accordingly, our proposed BHT-QAOA will provide broad opportunities to find all solutions for many classical Boolean problems constructed as Hamiltonians ($H_C$ and $H_M$), which are neither designed nor solved using the standard QAOA [1, 2]. Thus, various classical Boolean problems for the applications of digital logic circuits, robotics, and machine learning can be realized as Hamiltonians and then solved using BHT-QAOA in the hybrid classical-quantum domain.

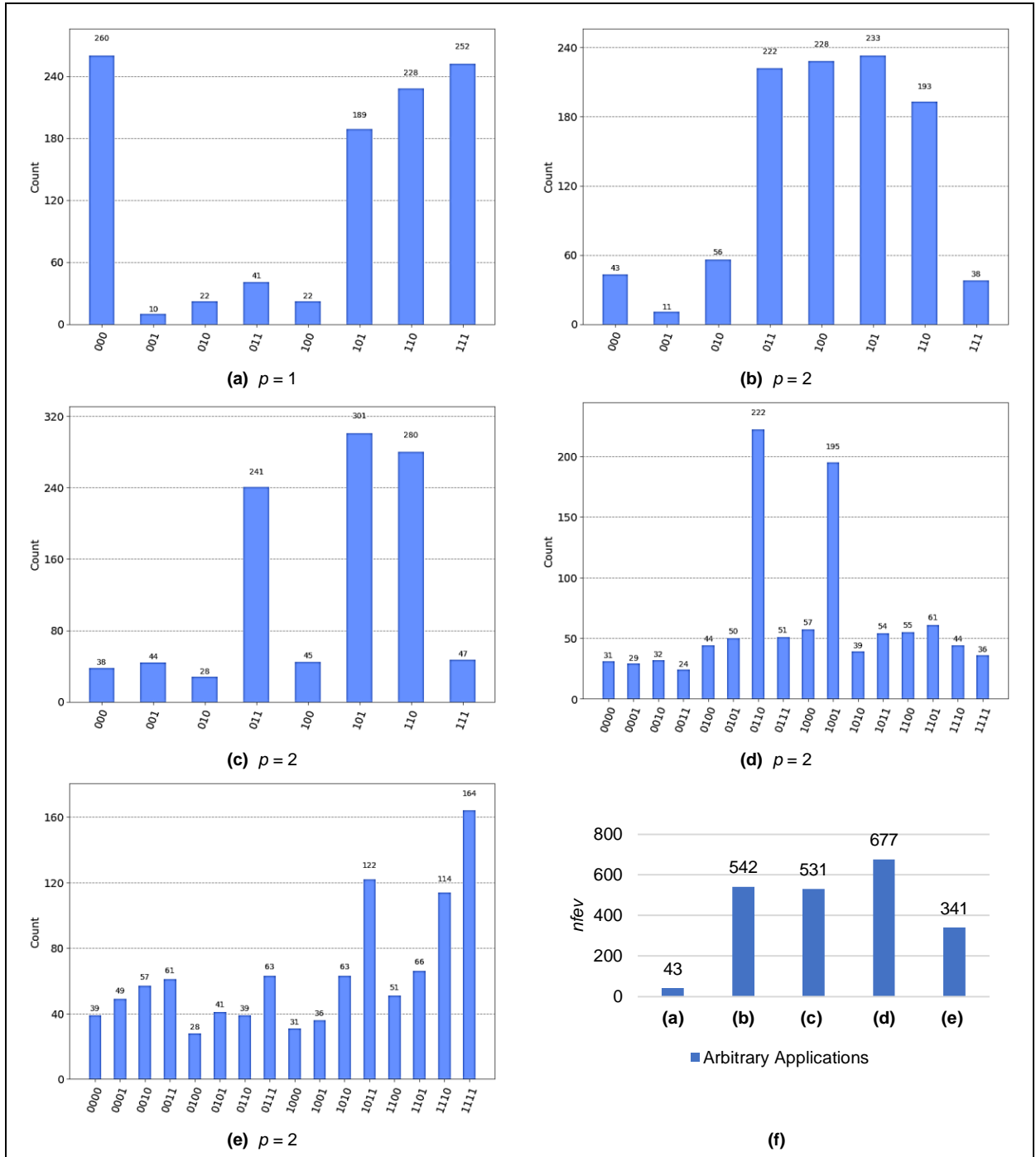

**Figure 6.** The final measured solutions for arbitrary applications executed once with `ibm_brisbane` QPU (for 1024 shots): **(a)** the solutions ['000', '101', '110', '111'] for the Boolean oracle in POS structure of Eq. (1) above, **(b)** the solutions ['011', '100', '101', '110'] for the Boolean oracle in SOP structure of Eq. (10) above, **(c)** the solutions ['011', '101', '110'] for the Boolean oracle in ESOP structure of Eq. (11) above, **(d)** the solutions ['0110', '1001'] for the 2×2 Sudoku game of Boolean oracle in CNF-XOR SAT structure of Eq. (12) above, **(e)** the solutions ['1011', '1110', '1111'] for the digital circuit of 4-bit conditioned half-adder for the Boolean oracle in mixed structures of Eq. (13) above, and **(f)** the required *nfev* for the SciPy optimization minimizer to successfully optimize the numerical values of *γ* and *β* for the above-mentioned applications. Note that, for the Boolean oracle of an application, (i) the first measured bit (the right-most or upper-right) of a solution is equivalent to the first input qubit, (ii) the last measured bit (the left-most or bottom-left) of a solution is equivalent to the last input qubit, (iii) *p* is the number of repetitions of Hamiltonians ($H_C$ and $H_M$) in a quantum circuit, (iv) *nfev* is the number of function evaluations, and (v) 'Count' is the total number of shots for all qubits.

# 4   Conclusion

A new methodology is proposed to solve arbitrary classical Boolean problems as Hamiltonians ($H_C$ and $H_M$) using the quantum approximate optimization algorithm (QAOA) [1, 2]. Our proposed methodology is termed the "Boolean-Hamiltonians Transform for QAOA (BHT-QAOA)". Our methodology of BHT-QAOA can be stated as follows: (i) an arbitrary classical Boolean problem is designed as a Boolean oracle in arbitrary structures, e.g., POS, SOP, ESOP, XOR SAT, just to name a few, (ii) this Boolean oracle (in any structure) is converted into its equivalent Boolean oracle in ESOP structure, unless it was firstly constructed in ESOP structure, (iii) this Boolean oracle in ESOP structure is transformed into its equivalent Phase oracle based on our modified set of Toffoli gates transformations, originally presented by Figgatt *et al.* [28], (iv) the Hamiltonians ($H_C$ and $H_M$) are generated from this transformed Phase oracle based on our modified set of Hamiltonian compositions, originally presented by Hadfield [38], and (v) all execution steps of the standard QAOA are performed using the above-generated $H_C$ and $H_M$, to find all approximated solutions for an arbitrary classical Boolean problem based on the optimized numerical values of $γ$ and $β$ for $H_C$ and $H_M$, respectively, from a classical optimization minimizer.

In BHT-QAOA, for an arbitrary classical Boolean problem, all ancilla qubits (including the output qubit) and the mirror (as the uncomputing part) of a quantum circuit will be completely removed, when transforming a Boolean oracle (in any structure) into its equivalent Phase oracle. In other words, (i) the total number of utilized qubits will be dramatically reduced to the number of input qubits only, and (ii) the total number of quantum gates will be significantly minimized for the final quantum circuit of a Phase oracle.

In this paper, arbitrary classical Boolean problems are constructed as Boolean oracles (in various structures), and then BHT-QAOA successfully finds all optimized approximated solutions for these problems using a classical optimization minimizer and an IBM quantum computer, since the standard QAOA and our proposed BHT-QAOA are considered as hybrid classical-quantum algorithms. In conclusion, further classical Boolean problems can be constructed as Boolean oracles (in mixed structures) for the practical engineering applications in the topics of digital synthesizers, computer vision, robotics, machine learning, just name a few, and our proposed BHT-QAOA will successfully solve such practical applications in the hybrid classical-quantum domain.